\documentstyle[11pt]{article}
\begin{document}

\title{\textbf{Interacting viscous entropy-corrected holographic scalar field models of dark energy
with time-varying $G$ in modified FRW cosmology}}

\author{F. Adabi$^{1}$ , K. Karami$^{2}$\thanks{E-mail: KKarami@uok.ac.ir} ,
F. Felegary$^{2}$, Z. Azarmi$^{2}$\\
$^{1}$\small{Department of Physics, Sanandaj Branch, Islamic Azad
University, Sanandaj, Iran}\\$^{2}$\small{Department of Physics,
University of Kurdistan, Pasdaran St., Sanandaj, Iran}}

\maketitle

\begin{abstract}
We study the entropy-corrected version of the holographic dark
energy (HDE) model in the framework of modified FRW cosmology. We
consider a non-flat universe filled with an interacting viscous
entropy-corrected HDE (ECHDE) with dark matter. We also include the
case of variable gravitational constant $G$ in our model. We obtain
the equation of state and the deceleration parameters of the
interacting viscous ECHDE. Moreover, we reconstruct the potential
and the dynamics of the quintessence, tachyon, K-essence and dilaton
scalar field models according to the evolutionary behavior of the
interacting viscous ECHDE model with time-varying $G$.
\end{abstract}


\clearpage
\section{Introduction}
The observed acceleration in the universe expansion rate
\cite{Riess} is usually attributed to the presence of an exotic kind
of energy, called dark energy (DE). Although the nature and
cosmological origin of DE is still enigmatic at present, a great
variety of models have been proposed to describe the DE (for review
see \cite{Padmanabhan,Copeland}).

Recently, a new DE candidate, namely, the holographic DE (HDE) based
on the holographic principle was proposed \cite{Suss1}. Following
Guberina et al. \cite{Guberina}, the HDE density can be derived from
the entropy bound. In the thermodynamics of the black hole
\cite{Bekenstein}, there is a maximum entropy in a box of size $L$,
namely, the Bekenstein-Hawking entropy bound $S_{\rm BH}\sim M_P^2
L^2$, which scales as the area of the box $A \sim L^2$, rather than
the volume $V \sim L^3$. Here $M_P$ is the reduced Planck Mass
$M_P^{-2}=8\pi G$. Also for a macroscopic system in which
self-gravitation effects can be disregarded, the Bekenstein entropy
bound $S_{\rm B}$ is given by the product of the energy
$E\sim\rho_{\Lambda}L^3$ and the length scale (IR cut-off) $L$ of
the system. Here $\rho_{\Lambda}$ is the quantum zero point energy
density caused by the UV cut-off $\Lambda$. Requiring $S_{\rm B}\leq
S_{\rm BH}$, namely $EL\leq M_P^2 L^2$, one has $\rho_{\Lambda}\leq
M_P^2L^{-2}$. If the largest cut-off $L$ is taken for saturating
this inequality, we get the HDE density as
\begin{equation}
\rho_{\Lambda}=3c^2M_P^2L^{-2},\label{rhoHDE}
\end{equation}
where $c$ is a numerical constant. Recent observational data, which
have been used to constrain the HDE model, show that for the
non-flat universe $c=0.815_{-0.139}^{+0.179}$ \cite{Li5}, and for
the flat case $c=0.818_{-0.097}^{+0.113}$ \cite {Li6}. Li \cite{Li}
showed that the cosmic coincidence problem can be resolved by
inflation in the HDE model, provided the minimal number of
e-foldings. The HDE models have been studied widely in the
literature \cite{Enqvist,Elizalde2,Guberina1,Guberina2,Karami1}.
Indeed, the definition and derivation of HDE density depends on the
entropy-area relationship $S_{\rm BH} = A/(4G)$, where $A\sim L^2$
is the area of horizon. However, this definition can be modified
from the inclusion of quantum effects, motivated from the loop
quantum gravity (LQG). These quantum corrections provided to the
entropy-area relationship leads to the curvature correction in the
Einstein-Hilbert action and vice versa \cite{Zhu}. The corrected
entropy takes the form \cite{modak}
\begin{equation}
S_{\rm
BH}=\frac{A}{4G}+\tilde{\alpha}\ln\left(\frac{A}{4G}\right)+\tilde{\beta},\label{entropy}
\end{equation}
where $\tilde{\alpha}$ and $\tilde{\beta}$ are dimensionless
constants of order unity. The exact values of these constants are
not yet determined and still an open issue in quantum gravity. These
corrections arise in the black hole entropy in LQG due to thermal
equilibrium fluctuations and quantum fluctuations \cite{Rovelli}.
Taking the corrected entropy-area relation (\ref{entropy}) into
account, and following the derivation of HDE (especially the one
shown in \cite{Guberina}), the HDE density will be modified as well.
On this basis, Wei \cite{HW} proposed
 the energy density of the so-called ``entropy-corrected HDE''
 (ECHDE) in the form
\begin{equation}
\rho_{\Lambda}=\frac{3c^2}{8\pi G
L^2}+\frac{\alpha}{L^4}\ln\left(\frac{L^2}{8\pi
G}\right)+\frac{\beta}{L^4},\label{ECHDE}
\end{equation}
where $\alpha$ and $\beta$ are dimensionless constants of order
unity. In the special case $\alpha=\beta=0$, the above equation
yields the well-known HDE density (\ref{rhoHDE}). Since the last two
terms in Eq. (\ref{ECHDE}) can be comparable to the first term only
when $L$ is very small, the corrections make sense only at the early
stage of the universe. When the universe becomes large, ECHDE
reduces to the ordinary HDE model. The ECHDE models have arisen a
lot of enthusiasm recently and have examined in ample detail by
\cite{Khodam,Karami2}.

Reconstructing the HDE scalar field models of DE is one of
interesting issue which has been investigated in the literature
\cite{Zhang11,Karami,Wu}. The HDE models are originated from some
considerations of the features of the quantum theory of gravity. On
the other hand, the scalar field models (such as quintessence,
tachyon, K-essence and dilaton) are often regarded as an effective
description of an underlying theory of DE \cite{Wu}. The scalar
field models can mimic cosmological constant at the present epoch
and can also alleviate the fine tuning and coincidence problems
\cite{Ali}. Therefore it becomes meaningful to reconstruct the
scalar field models from some DE models possessing some significant
features of the LQG theory, such as ECHDE models.

Here our aim is to investigate the correspondence between the
interacting viscous ECHDE and scalar field models of DE such as
quintessence, tachyon, K-essence and dilaton scalar fields and
obtain the evolutionary form of these fields with time-varying $G$.
There are significant indications that $G$ can by varying, being a
function of time or equivalently of the scale factor \cite{Inn}. In
particular, observations of Hulse-Taylor binary pulsar B$1913+16$
lead to the estimation $\dot{G}/G\sim2\pm4\times10^{-12}{\rm
yr}^{-1}$ \cite{Damour,kogan}, while helio-seismological data
provide the bound $-1.6\times10^{-12}{\rm yr}^{-1}<\dot{G}/G<0$
\cite{guenther}. Similarly, type Ia supernova observations  give the
best upper bound of the variation of $G$ as $-10^{-11} {\rm yr}^{-1}
\leq \frac{\dot G}{G}<0$ at redshifts $z \simeq 0.5$
\cite{Gaztanaga}, while astro-seismological data from the pulsating
white dwarf star G117-B15A lead to $\left|\frac{\dot G}{G}\right|
\leq 4.10 \times 10^{-11} {\rm yr}^{-1}$ \cite{Biesiada}.

This paper is organized as follows. In section 2, we investigate the
viscous ECHDE model with time-varying $G$ in the presence of
interaction with DM and in the framework of modified FRW cosmology.
In section 3, we establish a correspondence between the interacting
viscous ECHDE with time-varying $G$ and the quintessence, tachyon,
K-essence and dilaton scalar field models of DE. Section 4 is
devoted to our conclusions.

\section{Interacting viscous ECHDE with time-varying $G$ in modified FRW universe}

Within the framework of the FRW metric,
\begin{equation}
{\rm d}s^2=-{\rm d}t^2+a^2(t)\left(\frac{{\rm d}r^2}{1-kr^2}+r^2{\rm
d}\Omega^2\right),\label{metric}
\end{equation}
for the non-flat FRW universe containing the ECHDE and dark matter
(DM), the first modified Friedmann equation corresponding to the
corrected entropy-area relation (\ref{entropy}) is given by
\cite{Cai}
\begin{equation}
H^2+\frac{k}{a^2}+\frac{\tilde{\alpha}G}{2\pi}\left(H^2+\frac{k}{a^2}\right)^2=\frac{8\pi
G}{3}(\rho_{\rm m}+\rho_\Lambda),\label{friedmann}
\end{equation}
where $k=0, 1, -1$ represent a flat, closed and open FRW universe,
respectively. Also $\rho_\Lambda$ and $\rho_{\rm m}$ are the energy
density of ECHDE and DM, respectively. Using the following
definitions
\begin{equation}
\Omega_{\rm m}=\frac{8\pi G\rho_{\rm
m}}{3H^2},~~~~~\Omega_\Lambda=\frac{8\pi G \rho_\Lambda}{3H^2},~~~~~
\Omega_k=\frac{k}{a^2H^2},\label{omega123}
\end{equation}

\begin{equation}
\Omega_{\tilde{\alpha}}=\frac{\tilde{\alpha}G
H^2}{2\pi}(1+\Omega_k)^2,\label{friedmann2}
\end{equation}
one can rewrite the modified Friedmann equation (\ref{friedmann}) as
\begin{equation}
1+\Omega_k+\Omega_{\tilde{\alpha}}=\Omega_{\rm
m}+\Omega_\Lambda.\label{friedmann2}
\end{equation}
From definition $\rho_{\Lambda}=3H^2\Omega_{\Lambda}/(8\pi G)$ and
using Eq. (\ref{ECHDE}), we get
\begin{equation}
\Omega_\Lambda=\frac{c^2}{L^2 H^2}\gamma_c,\label{HL}
\end{equation}
where
\begin{equation}
\gamma_c=1+\frac{8\pi
G}{3c^2L^2}\left[\alpha\ln\left(\frac{L^2}{8\pi
G}\right)+\beta\right].\label{gamma}
\end{equation}
Note that to obtain an accelerating universe, following Huang and Li
\cite{Huang} the IR cut-off $L$ for a non-flat universe should be
defined as
\begin{equation}
L=a(t)\frac{\sin n\big(\sqrt{|k|}y\big)}{\sqrt{|k|}},\label{L}
\end{equation}
where
\begin{equation}
\frac{\sin n\big(\sqrt{|k|}y\big)}{\sqrt{|k|}}=\left\{
\begin{array}{ll}
\sin y,& k=1, \\
y,& k=0, \\
\sinh y,&k=-1, \\
\end{array}
\right.\label{sinn}
\end{equation}
and
\begin{equation}
y=\frac{R_{\rm h}}{a(t)}=\int_t^\infty \frac{{\rm d}
t}{a(t)}=\int_0^{r} \frac{{\rm d}r}{\sqrt{1-kr^2}}=\left\{
\begin{array}{ll}
\sin^{\rm -1}r,& k=1, \\
r,& k=0, \\
\sinh^{\rm -1}r,&k=-1. \\
\end{array}
\right.\label{y}
\end{equation}
Here $R_{\rm h}$ is the radial size of the event horizon measured in
the $r$ direction and $L$ is the radius of the event horizon
measured on the sphere of the horizon \cite{Huang}. For a flat
universe, $L=R_{\rm h}$.

Taking time derivative of Eq. (\ref{L}) and using (\ref{HL}) yields
\begin{equation}
\dot{L}=\left(\frac{c^2\gamma_c}{\Omega_{\Lambda}}\right)^{1/2}-\cos
n\Big(\sqrt{|k|}y\Big),\label{Ldot1}
\end{equation}
where
\begin{equation}
\cos n\Big(\sqrt{|k|}y\Big)=\left\{
\begin{array}{ll}
\cos y,& k=1, \\
1,& k=0, \\
\cosh y,&k=-1. \\
\end{array}
\right.\label{cosn1}
\end{equation}
Using Eqs. (\ref{omega123}), (\ref{HL}), (\ref{L}) and (\ref{sinn}),
one can rewrite Eq. (\ref{cosn1}) as
\begin{equation}
\cos
n\Big(\sqrt{|k|}y\Big)=\left[1-\Omega_k\left(\frac{c^2\gamma_c}{\Omega_{\Lambda}}\right)\right]^{1/2}.\label{cosn}
\end{equation}
Hence, Eq. (\ref{Ldot1}) yields
\begin{equation}
\dot{L}=\left(\frac{c^2\gamma_c}{\Omega_{\Lambda}}\right)^{1/2}\left[1-\left(\frac{\Omega_{\Lambda}}{c^2\gamma_c}
-\Omega_k\right)^{1/2}\right].\label{Ldot}
\end{equation}
Here, we would like to generalize our study to the case where the
ECHDE model has viscosity property. In an isotropic and homogeneous
modified FRW universe, the dissipative effects arise due to the
presence of bulk viscosity in cosmic fluids. DE with bulk viscosity
has a peculiar property to cause accelerated expansion of phantom
type in the late evolution of the universe \cite{Brevik}. It can
also alleviate several cosmological puzzles like age problem,
coincidence problem and phantom crossing. The energy-momentum tensor
of the viscous fluid is
\begin{equation}
T_{\mu\nu}=\rho_\Lambda u_\mu
u_\nu+\tilde{p}_\Lambda(g_{\mu\nu}+u_\mu u_\nu),\label{tensor}
\end{equation}
where $u_{\mu}$ is the four-velocity vector, $g_{\mu\nu}$ is the
background metric and
\begin{equation}
\tilde{p}_\Lambda=p_\Lambda-3H\xi,\label{preshur}
\end{equation}
is the effective pressure of DE and $ \xi $ is the bulk viscosity
coefficient \cite{Zimdahl}.

We further assume there is an interaction between viscous ECHDE and
DM. The recent observational evidence provided by the galaxy cluster
Abell A586 supports the interaction between DE and DM
\cite{Bertolami8}. In the presence of interaction, $\rho_{\Lambda}$
and $\rho_{\rm m}$ do not conserve separately and the energy
conservation equations for viscous ECHDE and DM are
\begin{equation}
\dot{\rho
}_\Lambda+3H\rho_\Lambda(1+\omega_\Lambda)=9H^2\xi-Q,\label{rho
dot2}
\end{equation}
\begin{equation}
\dot{\rho}_{\rm m}+3H\rho_{\rm m}=Q,\label{eqCDM}
\end{equation}
where $\omega_{\Lambda}=p_{\Lambda}/\rho_{\Lambda}$ is the equation
of state (EoS) parameter of the interacting viscous ECHDE and Q
stands for the interaction term. Following \cite{Pavon}, we shall
assume $Q=3b^2H(\rho_{\rm m}+\rho_{\Lambda})$ with the coupling
constant $b^2$. This expression for the interaction term was first
introduced in the study of the suitable coupling between a
quintessence scalar field and a pressureless cold DM field
\cite{Zimdahl,Amendola}. Using Eq. (\ref{friedmann2}), the
interaction term Q can be rewritten as
\begin{equation}
Q=3b^2H\rho_\Lambda\left(\frac{1+\Omega_k+\Omega_{\tilde{\alpha}}}{\Omega_{\Lambda}}\right).\label{Qequ}
\end{equation}
Taking time derivative of Eq. (\ref{ECHDE}) and using Eqs.
(\ref{HL}), (\ref{gamma}) and (\ref{Ldot}), one can obtain
\begin{equation}
\dot{\rho}_\Lambda=-2H\rho_\Lambda\left[2Y+\frac{1}{\gamma_c}\left(\frac{\acute{G}}{2G}-Y\right)\left(1+\frac{8\pi
G\alpha H^2\Omega_{\Lambda}}{3c^4\gamma_c}\right)
\right],\label{rhodot3}
\end{equation}
where
\begin{equation}
Y=1-\left(\frac{\Omega_{\Lambda}}{c^2\gamma_c}\right)^{1/2}\cos
n\Big(\sqrt{|k|}y\Big)=1-\left(\frac{\Omega_{\Lambda}}{c^2\gamma_c}
-\Omega_k\right)^{1/2},\label{psi}
\end{equation}
and the prime denotes the derivative with respect to $x=\ln a$.

Substituting Eqs. (\ref{Qequ}) and (\ref{rhodot3}) in (\ref{rho
dot2}), using Eq. (\ref{HL}) and assuming $\xi=\varepsilon
H^{-1}\rho_\Lambda$ \cite{Sheykhi1} where $\varepsilon $ is a
constant parameter, then one can obtain the EoS parameter of the
interacting viscous ECHDE as
\begin{eqnarray}
\omega_\Lambda=
-1+3\varepsilon+\frac{4}{3}Y+\frac{2}{3\gamma_c}\left(\frac{\acute{G}}{2G}-Y\right)\left(1+\frac{8\pi
G\alpha H^2\Omega_{\Lambda}}{3c^4\gamma_c}\right)
-b^2\left(\frac{1+\Omega_k+\Omega_{\tilde{\alpha}}}{\Omega_{\Lambda}}\right).\label{eos}
\end{eqnarray}
If we set $\tilde{\alpha}=0=\Omega_{\tilde{\alpha}}$ and
$\varepsilon=\acute{G}=0$, then Eq. (\ref{eos}) reduces to the EoS
parameter of interacting ECHDE in Einstein gravity \cite{ Karami2}
\begin{equation}
\omega_\Lambda=-1-\frac{2Y}{3\gamma_c} \left(1-2\gamma_c+\frac{8\pi
G\alpha
H^2\Omega_{\Lambda}}{3c^4\gamma_c}\right)-b^2\left(\frac{1+\Omega_k}{\Omega_\Lambda}\right).\label{eos2}
\end{equation}
Also in the absence of correction terms ($ \alpha=\beta=0 $), from
Eq. (\ref{gamma}) we have $\gamma_c=1$ and Eq. (\ref{eos2}) recovers
the EoS parameter of interacting HDE in Einstein gravity \cite{wang}
\begin{equation}
\omega_\Lambda=-\frac{1}{3}-\frac{2}{3c}\Omega_{\Lambda}^{1/2}\cos
n\Big(\sqrt{|k|}y\Big)-b^2\left(\frac{1+\Omega_k}{\Omega_\Lambda}\right).\label{eos3}
\end{equation}
Note that as we already mentioned, at the very early stage when the
universe undergoes an inflation phase, the correction terms in the
ECHDE density (\ref{ECHDE}) become important. After the end of the
inflationary phase, the universe subsequently enters in the
radiation and then matter dominated eras. In these two epochs, since
the universe is much larger, the entropy-corrected terms to ECHDE,
namely the last two terms in Eq. (\ref{ECHDE}), can be safely
ignored. During the early inflation era when the correction terms
make sense in the ECHDE density (\ref{ECHDE}), the Hubble parameter
$H$ is constant and $a=a_0e^{Ht}$. Hence the Hubble horizon $H^{-1}$
and the future event horizon $R_{\rm h}=a\int_t^\infty \frac{{\rm d}
t}{a}$ will coincide i.e. $R_{\rm h} = H^{-1}=$ const. On the other
hand, since an early inflation era leads to a flat universe, we have
$L = R_{\rm h} = H^{-1}=$ const. Also from Eqs. (\ref{HL}) and
(\ref{cosn}) we have $\frac{\Omega_{\Lambda}}{c^2\gamma_c}=1$ and
$\cos n\Big(\sqrt{|k|}y\Big)=1$, hence Eq. (\ref{psi}) gives $Y=0$.
Therefore, during the early inflation era, Eq. (\ref{eos}) reduces
to
\begin{equation}
\omega_\Lambda=
-1+3\varepsilon+\frac{c^2}{3\Omega_{\Lambda}}\frac{\acute{G}}{G}\left(1+\frac{8\pi
G\alpha H^2}{3c^2}\right)
-\frac{b^2}{\Omega_{\Lambda}}\left(1+\frac{\tilde{\alpha}
GH^2}{2\pi}\right),\label{eosinf}
\end{equation}
where
\begin{equation}
\Omega_{\Lambda}=c^2\gamma_c=c^2+\frac{8\pi
GH^2}{3}\left[\alpha\ln\left(\frac{1}{8\pi
GH^2}\right)+\beta\right].
\end{equation}
Taking time derivative of Eq. (\ref{HL}) and using
$\dot{\Omega}_\Lambda=H\Omega^{'}_{\Lambda}$, one can get the
equation of motion for $\Omega_\Lambda$ as
\begin{equation}
\Omega^{'}_{\Lambda}=-2\Omega_\Lambda
\left(\frac{\dot{H}}{H^2}+\frac{\dot{L}}{LH}-\frac{\dot{\gamma_c}}{2H\gamma_c}\right),\label{omega
perim1}
\end{equation}
where
\begin{eqnarray}
\frac{\dot{H}}{H^2}=\left(\frac{1+\Omega_k}{1+\Omega_k+2\Omega_{\tilde{\alpha}}}\right)
\left\{\frac{3}{2}(b^2-1)\Omega_{\rm
m}+\frac{\acute{G}}{2G}(1+\Omega_k)\right.
~~~~~~~~~~~~~~~~~~~~~~~~~~~~~~~~~\nonumber\\\left.
+\left[\frac{3}{2}b^2-2Y-\frac{1}{\gamma_c}\left(\frac{\acute{G}}{2G}-Y\right)
\left(1+\frac{8\pi G\alpha
H^2\Omega_{\Lambda}}{3c^4\gamma_c}\right)\right]\Omega_{\Lambda}\right\}+\Omega_k,\label{Hd}
\end{eqnarray}
and
\begin{equation}
-\frac{\dot{\gamma}_c}{2H\gamma_c}=\frac{1}{\gamma_c}\left(\frac{\acute{G}}{2G}-Y\right)
\left(1-\gamma_c+\frac{8\pi G\alpha
H^2\Omega_{\Lambda}}{3c^4\gamma_c}\right).\label{gammad}
\end{equation}
Using Eqs. (\ref{HL}), (\ref{Ldot}), (\ref{Hd}) and (\ref{gammad}),
one can rewrite (\ref{omega perim1}) as
\begin{eqnarray}\label{omega perim2}
\frac{\Omega_{\Lambda}^{'}}{\Omega_\Lambda}&=&-2\left(\frac{1+\Omega_k}{1+\Omega_k+2\Omega_{\tilde{\alpha}}}\right)
\left(\frac{3}{2}(b^2-1)\Omega_{\rm
m}+\frac{\acute{G}}{2G}(1+\Omega_k)+\frac{3}{2}b^2\Omega_{\Lambda}\right)
\nonumber\\
&&-2\left(
1-\frac{(1+\Omega_k)\Omega_{\Lambda}}{1+\Omega_k+2\Omega_{\tilde{\alpha}}}\right)\left[2Y+\frac{1}{\gamma_c}\left(\frac{\acute{G}}{2G}-Y\right)
\left(1+\frac{8\pi G\alpha
H^2\Omega_{\Lambda}}{3c^4\gamma_c}\right)\right]\nonumber\\
&&+\frac{\acute{G}}{G}-2\Omega_k .
\end{eqnarray}
For completeness, we give the deceleration parameter
\begin{equation}
q=-\frac{\ddot{a}}{a H^2}=-1-\frac{\dot{H}}{H^2}.\label{decelration}
\end{equation}
When deceleration parameter is combined with the Hubble parameter
and the dimensionless density parameters, form a set of useful
parameters for the description of the astrophysical observations.
Substituting Eq. (\ref{Hd}) in (\ref{decelration}) gives
\begin{eqnarray}
q=\left(\frac{1+\Omega_k}{1+\Omega_k+2\Omega_{\tilde{\alpha}}}\right)
\left\{\frac{3}{2}\Omega_{\rm
m}-\frac{3}{2}b^2(1+\Omega_k+\Omega_{\tilde{\alpha}})-\frac{\acute{G}}{2G}(1+\Omega_k)\right.
~~~~\nonumber\\\left.
+\left[2Y+\frac{1}{\gamma_c}\left(\frac{\acute{G}}{2G}-Y\right)
\left(1+\frac{8\pi G\alpha
H^2\Omega_{\Lambda}}{3c^4\gamma_c}\right)\right]\Omega_{\Lambda}\right\}-(1+\Omega_k).\label{qequ}
\end{eqnarray}
If we set $\tilde{\alpha}=0=\Omega_{\tilde{\alpha}}$,
$\varepsilon=\acute{G}=0$ and $\alpha=\beta=0$ then from Eq.
(\ref{gamma}) we have $\gamma_c=1$ and Eq. (\ref{qequ}) reduces to
the deceleration parameter of interacting HDE in Einstein gravity
\cite{wang}
\begin{equation}
q = - \frac{\Omega_\Lambda}{2}+
\frac{1}{2}(1-3b^2)(1+\Omega_k)-\frac{\Omega_\Lambda^{3/2}}{c}\cos
n\Big(\sqrt{|k|}y\Big).\label{q3}
\end{equation}

\section{Correspondence between the interacting viscous ECHDE and scalar field models of
DE}

Here, we suggest a correspondence between the interacting viscous
ECHDE model with the quintessence, tachyon, K-essence and dilaton
scalar field models in the context of the modified FRW cosmology. To
establish this correspondence, we compare the ECHDE density
(\ref{ECHDE}) with the corresponding scalar field model density and
also equate the equations of state for this models with the EoS
parameter given by (\ref{eos}).


\subsection{Interacting viscous ECHDE quintessence model}

The quintessence scalar filed model of DE was proposed to justify
the late-time acceleration of the universe. For the quintessence
scalar field $\phi$, the energy density and pressure are given by
\cite{Copeland}
\begin{equation}
\rho_Q=\frac{1}{2}\dot \phi^2+V(\phi),\label{ro q}
\end{equation}
\begin{equation}
p_Q=\frac{1}{2}\dot \phi^2-V(\phi),\label{p q}
\end{equation}
and the EoS parameter is obtained as
\begin{equation}
\omega_Q=\frac{p_Q}{\rho_Q}=\frac{\dot \phi^2-2V(\phi)}{\dot
\phi^2+2V(\phi)}.\label{w q}
\end{equation}
Equating Eq. (\ref{w q}) with the EoS parameter (\ref{eos}),
$\omega_Q=\omega_\Lambda$, and also equating Eq. (\ref{ro q}) with
(\ref{ECHDE}), $\rho_Q=\rho_\Lambda$, we have
\begin{equation} \dot
\phi^2=(1+\omega_\Lambda)\rho_\Lambda,\label{phidot2-2}
\end{equation}
\begin{equation}
V(\phi)=\frac{1}{2}(1-\omega_\Lambda)\rho_\Lambda.\label{Vphi-2}
\end{equation}
Substituting Eqs. (\ref{ECHDE}) and (\ref{eos}) into Eqs.
(\ref{phidot2-2}) and (\ref{Vphi-2}), the kinetic energy term and
the potential energy of the quintessence scalar filed are obtained
as follows
\begin{equation}
\dot\phi^2=\frac{3H^2\Omega_\Lambda}{8\pi
G}\left[3\varepsilon+\frac{4}{3}Y+\frac{2}{3\gamma_c}\left(\frac{\acute{G}}{2G}-Y\right)\left(1+\frac{8\pi
G\alpha H^2\Omega_{\Lambda}}{3c^4\gamma_c}\right)
-b^2\left(\frac{1+\Omega_k+\Omega_{\tilde{\alpha}}}{\Omega_{\Lambda}}\right)\right],\label{fi
dot q}
\end{equation}
\begin{eqnarray}
V(\phi)=\frac{3H^2\Omega_\Lambda}{16\pi
G}\left[2-3\varepsilon-\frac{4}{3}Y-\frac{2}{3\gamma_c}\left(\frac{\acute{G}}{2G}-Y\right)\left(1+\frac{8\pi
G\alpha H^2\Omega_{\Lambda}}{3c^4\gamma_c}\right)
\right.\nonumber\\\left.+b^2\left(\frac{1+\Omega_k+\Omega_{\tilde{\alpha}}}{\Omega_{\Lambda}}\right)\right].\label{pot
q}
\end{eqnarray}
Integrating Eq. (\ref{fi dot q}) with respect to the scale factor
$a$ yields the evolutionary form of the quintessence scalar field as
\begin{eqnarray}
\phi(a)-\phi(a_0)=\int^{a}_{a_0}\frac{{\rm
d}a}{a}\left\{\frac{3\Omega_\Lambda}{8\pi
G}\left[3\varepsilon+\frac{4}{3}Y+\frac{2}{3\gamma_c}\left(\frac{\acute{G}}{2G}-Y\right)\left(1+\frac{8\pi
G\alpha H^2\Omega_{\Lambda}}{3c^4\gamma_c}\right)
\right.\right.\nonumber\\\left.\left.-b^2\left(\frac{1+\Omega_k+\Omega_{\tilde{\alpha}}}{\Omega_{\Lambda}}\right)\right]\right\}^{1/2},
\end{eqnarray}
where $a_0$ is the scale factor at the present time.
\subsection{Interacting viscous ECHDE tachyon
model}

It has been suggested that the tachyon scalar field model, in a
class of string theories, can act as a source of DE depending upon
the form of the tachyon potential. A rolling tachyon has an
interesting EoS whose parameter smoothly interpolates between $-1$
and $0$ \cite{gibbons}. This has led to a flurry of attempts being
made to construct viable cosmological models using the tachyon as a
suitable candidate for the inflation at high energy. For the
effective Lagrangian density of the tachyon field as \cite{Sen}
\begin{equation}
{\mathcal{L}}=-V(\phi)\sqrt{1+\partial_{\mu}\phi
\partial^{\mu}\phi},
\end{equation}
the energy density and pressure are given by \cite{Sen}
\begin{equation}
\rho_{T}=\frac{V(\phi)}{\sqrt{1-\dot{\phi}^{2}}},\label{rhot}
\end{equation}
\begin{equation}
p_{T}=-V(\phi)\sqrt{1-\dot{\phi}^{2}},
\end{equation}
where $V(\phi)$ is the tachyon potential. The EoS parameter of the
tachyon scalar field is obtained as
\begin{equation}
\omega_{T}=\frac{p_{T}}{\rho_{T}}=\dot{\phi}^{2}-1.\label{wt}
\end{equation}
If we establish the correspondence between the interacting viscous
ECHDE and tachyon DE, then equating Eq. (\ref{wt}) with (\ref{eos}),
$\omega_T=\omega_\Lambda$, and also equating Eq. (\ref{rhot}) with
(\ref{ECHDE}), $\rho_T=\rho_\Lambda$, we obtain
\begin{eqnarray}
\dot{\phi}^{2}=3\varepsilon+\frac{4}{3}Y+\frac{2}{3\gamma_c}\left(\frac{\acute{G}}{2G}-Y\right)\left(1+\frac{8\pi
G\alpha H^2\Omega_{\Lambda}}{3c^4\gamma_c}\right)
-b^2\left(\frac{1+\Omega_k+\Omega_{\tilde{\alpha}}}{\Omega_{\Lambda}}\right),\label{fi
dotT}
\end{eqnarray}
\begin{eqnarray}
V(\phi)=\frac{3H^2\Omega_\Lambda}{8\pi
G}\left[1-3\varepsilon-\frac{4}{3}Y-\frac{2}{3\gamma_c}\left(\frac{\acute{G}}{2G}-Y\right)\left(1+\frac{8\pi
G\alpha H^2\Omega_{\Lambda}}{3c^4\gamma_c}\right)
\right.\nonumber\\\left.+b^2\left(\frac{1+\Omega_k+\Omega_{\tilde{\alpha}}}{\Omega_{\Lambda}}\right)\right]^{1/2}.\label{pot
T}
\end{eqnarray}
From Eq. (\ref{fi dotT}), the evolutionary form of the tachyon
scalar field is obtained as
\begin{eqnarray}
\phi(a)-\phi(a_0)=\int_{a_0}^{a}\frac{{\rm
d}a}{Ha}\left[3\varepsilon+\frac{4}{3}Y+\frac{2}{3\gamma_c}\left(\frac{\acute{G}}{2G}-Y\right)\left(1+\frac{8\pi
G\alpha H^2\Omega_{\Lambda}}{3c^4\gamma_c}\right)
\right.\nonumber\\\left.-b^2\left(\frac{1+\Omega_k+\Omega_{\tilde{\alpha}}}{\Omega_{\Lambda}}\right)\right]^{1/2}.
\end{eqnarray}
\subsection{Interacting viscous ECHDE
K-essence model}

The scalar field model known as K-essence is also used to explain
the observed late-time acceleration of the universe. It is well
known that K-essence scenarios have attractor-like dynamics, and
therefore avoid the fine-tuning of the initial conditions for the
scalar field. K-essence is characterized by a scalar field with a
non-canonical kinetic energy. The scalar field action of the
K-essence is a function of $\phi$ and $\chi=\dot{\phi}^2/2$ as
\cite{Chiba, Picon3}
\begin{equation} S=\int {\rm d} ^{4}x\sqrt{-{\rm
g}}~p(\phi,\chi),
\end{equation}
where
\begin{equation}
p(\phi,\chi)=f(\phi)(-\chi+\chi^{2}),
\end{equation}
is the pressure density of the K-essence field. Also the energy
density of the field $\phi$ is given by
\begin{equation}
\rho(\phi,\chi)=f(\phi)(-\chi+3\chi^{2}).\label{rhok}
\end{equation}
The EoS parameter of the K-essence scalar field is obtained as
\begin{equation}
\omega_{K}=\frac{p(\phi,\chi)}{\rho(\phi,\chi)}=\frac{\chi-1}{3\chi-1}.\label{wk}
\end{equation}
Equating Eq. (\ref{wk}) with the EoS parameter (\ref{eos}),
$\omega_{K}=\omega_{\Lambda}$, yields the solution for $\chi$ as
\begin{equation}
\chi=\frac{2-3\varepsilon-\frac{4}{3}Y-\frac{2}{3\gamma_c}\Big(\frac{\acute{G}}{2G}-Y\Big)\Big(1+\frac{8\pi
G\alpha H^2\Omega_{\Lambda}}{3c^4\gamma_c}\Big)
+b^2\Big(\frac{1+\Omega_k+\Omega_{\tilde{\alpha}}}{\Omega_{\Lambda}}\Big)}{4-9\varepsilon
-4Y-\frac{2}{\gamma_c}\Big(\frac{\acute{G}}{2G}-Y\Big)\Big(1+\frac{8\pi
G\alpha H^2\Omega_{\Lambda}}{3c^4\gamma_c}\Big)
+3b^2\Big(\frac{1+\Omega_k+\Omega_{\tilde{\alpha}}}{\Omega_{\Lambda}}\Big)}.\label{khi}
\end{equation}
Using $\dot{\phi}^2=2\chi$ and (\ref{khi}), one can get the
evolutionary form of the K-essence scalar field as
\begin{equation}
\phi(a)-\phi(a_0)=\int_{a_0}^{a}\frac{{\rm
d}a}{Ha}\left[\frac{4-6\varepsilon-\frac{8}{3}Y-\frac{4}{3\gamma_c}\Big(\frac{\acute{G}}{2G}-Y\Big)\Big(1+\frac{8\pi
G\alpha H^2\Omega_{\Lambda}}{3c^4\gamma_c}\Big)
+2b^2\Big(\frac{1+\Omega_k+\Omega_{\tilde{\alpha}}}{\Omega_{\Lambda}}\Big)}{4-9\varepsilon
-4Y-\frac{2}{\gamma_c}\Big(\frac{\acute{G}}{2G}-Y\Big)\Big(1+\frac{8\pi
G\alpha H^2\Omega_{\Lambda}}{3c^4\gamma_c}\Big)
+3b^2\Big(\frac{1+\Omega_k+\Omega_{\tilde{\alpha}}}{\Omega_{\Lambda}}\Big)}\right]^{1/2}.\label{K
F}
\end{equation}

\subsection{Interacting viscous ECHDE dilaton
model}

The dilaton scalar field model is also an interesting attempt to
explain the origin of DE using string theory. This model appears
from a four-dimensional effective low-energy string action and
includes higher-order kinetic corrections to the tree-level action
in low energy effective string theory. For the dilaton DE model, the
pressure and energy densities are given by \cite{Gasperini}
\begin{equation}
p_{D}=-\chi+c'e^{\lambda\phi}\chi^{2},
\end{equation}
\begin{equation}
\rho_{D}=-\chi+3c'e^{\lambda\phi}\chi^{2},\label{rhod}
\end{equation}
where $c'$ and $\lambda$ are positive constants and
$\chi=\dot{\phi}^2/2$. The EoS parameter of the dilaton scalar field
is obtained as
\begin{equation}
\omega_{D}=\frac{p_D}{\rho_D}=\frac{-1+c'e^{\lambda\phi}\chi}{-1+3c'e^{\lambda\phi}\chi}.\label{wd}
\end{equation}
Equating Eq. (\ref{wd}) with the EoS parameter (\ref{eos}),
$\omega_D=\omega_\Lambda$, we find the following solution
\begin{equation}
c'e^{\lambda\phi}\chi=\frac{2-3\varepsilon-\frac{4}{3}Y-\frac{2}{3\gamma_c}\left(\frac{\acute{G}}{2G}-Y\right)\left(1+\frac{8\pi
G\alpha H^2\Omega_{\Lambda}}{3c^4\gamma_c}\right)
+b^2\left(\frac{1+\Omega_k+\Omega_{\tilde{\alpha}}}{\Omega_{\Lambda}}\right)}{4-9\varepsilon
-4Y-\frac{2}{\gamma_c}\left(\frac{\acute{G}}{2G}-Y\right)\left(1+\frac{8\pi
G\alpha H^2\Omega_{\Lambda}}{3c^4\gamma_c}\right)
+3b^2\left(\frac{1+\Omega_k+\Omega_{\tilde{\alpha}}}{\Omega_{\Lambda}}\right)},\label{khi
D}
\end{equation}
then using $\dot{\phi}^2=2\chi$, we obtain
\begin{equation}
e^{\frac{\lambda\phi}{2}}\dot{\phi}=\frac{1}{\sqrt{c'}}\left[\frac{4-6\varepsilon-\frac{8}{3}Y-\frac{4}{3\gamma_c}\left(\frac{\acute{G}}{2G}-Y\right)\left(1+\frac{8\pi
G\alpha H^2\Omega_{\Lambda}}{3c^4\gamma_c}\right)
+2b^2\left(\frac{1+\Omega_k+\Omega_{\tilde{\alpha}}}{\Omega_{\Lambda}}\right)}{4-9\varepsilon
-4Y-\frac{2}{\gamma_c}\left(\frac{\acute{G}}{2G}-Y\right)\left(1+\frac{8\pi
G\alpha H^2\Omega_{\Lambda}}{3c^4\gamma_c}\right)
+3b^2\left(\frac{1+\Omega_k+\Omega_{\tilde{\alpha}}}{\Omega_{\Lambda}}\right)}\right]^{1/2}.
\end{equation}
Integrating with respect to $a$, we get
\begin{eqnarray}
e^\frac{\lambda\phi(a)}{2}=e^\frac{\lambda\phi(a_0)}{2}
~~~~~~~~~~~~~~~~~~~~~~~~~~~~~~~~~~~~~~~~~~~~~~~~~~~~~~~~~~~~~~~~~~~~~~~~~~~~~~~~~~~~~~~~~~~\nonumber\\
+\frac{\lambda}{2\sqrt{c'}}\int_{a_0}^{a}\frac{{\rm
d}a}{Ha}\left[\frac{4-6\varepsilon-\frac{8}{3}Y-\frac{4}{3\gamma_c}\left(\frac{\acute{G}}{2G}-Y\right)\left(1+\frac{8\pi
G\alpha H^2\Omega_{\Lambda}}{3c^4\gamma_c}\right)
+2b^2\left(\frac{1+\Omega_k+\Omega_{\tilde{\alpha}}}{\Omega_{\Lambda}}\right)}{4-9\varepsilon
-4Y-\frac{2}{\gamma_c}\left(\frac{\acute{G}}{2G}-Y\right)\left(1+\frac{8\pi
G\alpha H^2\Omega_{\Lambda}}{3c^4\gamma_c}\right)
+3b^2\left(\frac{1+\Omega_k+\Omega_{\tilde{\alpha}}}{\Omega_{\Lambda}}\right)}\right]^{1/2}.
\end{eqnarray}
Finally, the evolutionary form of the dilaton scalar field is
obtained as
\begin{eqnarray}
\phi(a)=\frac{2}{\lambda}\ln\left\{e^\frac{\lambda\phi(a_0)}{2}\right.
~~~~~~~~~~~~~~~~~~~~~~~~~~~~~~~~~~~~~~~~~~~~~~~~~~~~~~~~~~~~~~~~~~~~~~~~~~~~~~~~~~~~~~~~~~\nonumber\\\left.
+\frac{\lambda}{2\sqrt{c'}}\int_{a_0}^{a}\frac{{\rm
d}a}{Ha}\left[\frac{4-6\varepsilon-\frac{8}{3}Y-\frac{4}{3\gamma_c}\left(\frac{\acute{G}}{2G}-Y\right)\left(1+\frac{8\pi
G\alpha H^2\Omega_{\Lambda}}{3c^4\gamma_c}\right)
+2b^2\left(\frac{1+\Omega_k+\Omega_{\tilde{\alpha}}}{\Omega_{\Lambda}}\right)}{4-9\varepsilon
-4Y-\frac{2}{\gamma_c}\left(\frac{\acute{G}}{2G}-Y\right)\left(1+\frac{8\pi
G\alpha H^2\Omega_{\Lambda}}{3c^4\gamma_c}\right)
+3b^2\left(\frac{1+\Omega_k+\Omega_{\tilde{\alpha}}}{\Omega_{\Lambda}}\right)}\right]^{1/2}\right\}.
\end{eqnarray}

\section{Conclusions}
Among various candidates to explain the cosmic accelerated
expansion, only the HDE model is based on the entropy-area relation.
Note that the entropy-area relation depends on the gravity theory.
When applying the curvature corrections to the gravity theory, it
yields quantum corrections to the entropy-area relation. The ECHDE
density is obtained by adding the correction terms to the HDE
density which is motivated from the LQG.

Here, we investigated the entropy-corrected version of the HDE in
the framework of modified FRW cosmology. We considered a spatially
non-flat universe filled with the interacting viscous ECHDE with DM.
Then, we extended our study to the case where the gravitational
constant $G$ varies with time. A time-varying $G$ has some
theoretical advantages such as alleviating the DM problem
\cite{Goldman}, the cosmic coincidence problem \cite{Jamil} and the
discrepancies in Hubble parameter value \cite{Ber}. We derived an
exact differential equation that determines the evolution of the
ECHDE density parameter. Furthermore, we obtained the EoS and
deceleration parameters of the interacting viscous ECHDE model.

We also established a correspondence between the interacting viscous
ECHDE model and the quintessence, tachyon, K-essence and dilaton
scalar field models of DE in the modified FRW scenario including
time-varying $G$. We adopted the viewpoint that these scalar field
models of DE are effective theories of an underlying theory of DE.
Thus, we should be capable of using these scalar field models to
mimic the evolving behavior of the ECHDE and reconstructing the
scalar field models according to the evolutionary behavior of the
ECHDE. Finally, we reconstructed the potentials and the dynamics of
these scalar field models which describe quintessence, tachyon,
K-essence and dilaton cosmology.

We hope that the future high precision astronomical observations
like the type Ia supernovae (SNeIa) surveys, the shift parameter of
the cosmic microwave background (CMB) given by the Wilkinson
Microwave Anisotropy Probe (WMAP) observations, and the baryon
acoustic oscillation (BAO) measurement from the Sloan Digital Sky
Survey (SDSS) may be capable of determining the fine property of the
interacting viscous ECHDE model in modified FRW cosmology and
consequently reveal some significant features of the underlying
theory of DE.
\\
\\
\noindent{\textbf{Acknowledgements}\\
This work has been supported financially by Department of Physics,
Sanandaj Branch, Islamic Azad University, Sanandaj, Iran.


\end{document}